\documentclass[12pt,preprint]{aastex}

\usepackage{multirow,color}
\usepackage{bbding}
\usepackage{amssymb}
\usepackage{ulem}
\usepackage{color}
\usepackage{longtable}
\usepackage{url}
\usepackage[displaymath]{lineno}

\def\gsim{\;\lower4pt\hbox{${\buildrel\displaystyle >\over\sim}$}\;}
\def\lsim{\;\lower4pt\hbox{${\buildrel\displaystyle <\over\sim}$}\;}
\def\grls{\;\lower4pt\hbox{${\buildrel\displaystyle >\over <}$}\;}

\begin{document}

\title{Why is a flare-rich active region CME-poor?}

\author{Lijuan Liu,$^{1,4}$ Yuming Wang,$^{1,2}$ Jingxiu Wang,$^3$ Chenglong Shen,$^{1,2}$  Pinzhong Ye,$^{1}$ Rui Liu,$^{1,4}$ Jun Chen,$^{1,5}$ Quanhao Zhang,$^{1,5}$ and S. Wang$^{1,4}$}

\altaffiltext{1}{CAS Key Laboratory of Geospace Environment,
Department of Geophysics and Planetary Sciences, University of Science and
Technology of China, Hefei, Anhui 230026, China, (ymwang@ustc.edu.cn, ljliu@mail.ustc.edu.cn)}
\altaffiltext{2}{Synergetic Innovation Center of Quantum Information \& Quantum Physics, University of Science and Technology of China, Hefei, Anhui 230026, China}
\altaffiltext{3}{National Astronomical Observatories, Chinese Academy of Sciences, Beijing 100012, China}
\altaffiltext{4}{Collaborative Innovation Center of Astronautical Science and Technology, China}
\altaffiltext{5}{Mengcheng National Geophysical Observatory, University of Science and Technology of China}

\begin{abstract}
Solar active regions (ARs) are the major sources of two kinds of
the most violent solar eruptions, namely flares and coronal mass
ejections (CMEs). The largest AR in the past 24 years, NOAA AR
12192, crossed the visible disk from 2014 October 17 to 30, unusually 
produced more than one hundred flares, including 32 M-class and 6 X-class ones,
but only one small CME. Flares and CMEs are believed to be two
phenomena in the same eruptive process. Why is such a flare-rich AR
so CME-poor? We compared this AR with other four ARs; two were productive in both and two were inert. The investigation of the photospheric parameters based on the SDO/HMI vector magnetogram 
reveals that the flare-rich AR 12192, as the other two productive ARs, has larger magnetic flux, current and free magnetic energy than 
the two inert ARs, but contrast to the two productive ARs, it has no strong, concentrated current helicity along both sides 
of the flaring neutral line,
indicating the absence of a mature magnetic structure consisting of highly sheared or twisted field lines. 
Furthermore, the decay index above the AR 12192 is relatively low, showing strong constraint. 
These results suggest that productive ARs are always large and have enough current and free energy to power flares, but whether or not 
a flare is accompanied by a CME is seemingly related to (1) if there is mature sheared or twisted core field serving as the seed of the 
CME, (2) if the constraint of the overlying arcades is weak enough.

\end{abstract}

\section{Introduction} \label{sec:intro}
Both solar flares and coronal mass ejections (CMEs) indicate the rapid
release of a huge amount of magnetic energy in the solar corona, and in
particular, CMEs are the most important driving source of hazardous space weather near the
geospace. As the major producer of flares and CMEs,
active regions (ARs) have been studied for decades. It was revealed
based on lots of observational studies that parameters characterizing the AR's non-potentiality, e.g., shear
length, magnetic gradient, total electric current, free energy, are all correlated with the flare and CME productivity \citep[e.g.,][]{Canfield_etal_1999, Sammis_etal_2000, Falconer_etal_2002, Falconer_etal_2006,Leka_Barnes_2003a, Leka_Barnes_2003b, Jing_etal_2006, Ternullo_etal_2006, Schrijver_2007, Georgoulis_Rust_2007, Guo_etal_2007, Wang_Zhang_2008}, and particularly larger ARs are more likely to produce eruptions \citep[e.g.,][]{Tian_etal_2002a, Chen_etal_2011}. However, not all large ARs have similar productivities in flares and CMEs, some may be productive in flares only \citep[e.g.,][]{Tian_etal_2002a, Akiyama_etal_2007, ChenA_Wang_2012}. How to distinguish the productivity of an AR is a key issue in space weather forecasting, and still unsolved so far.

The recent super AR, 12192, crossing the visible solar disk during
2014 October 17 -- 30, caught a wide attention \citep{RHESSI_Nugget_239, Sun_etal_2015, Thalmann_etal_2015}.
It is the largest AR
since 1990 November, but produced only one small CME though a total
of 127 C-class and intenser flares, including 32 M-class and 6
X-class ones, were generated. Flares and CMEs are thought to be the consequences
of the same eruptive process \citep[e.g.,][]{Harrison_1995,
Lin_Forbes_2000}. Although the released energy of them during a
strong eruption are on the same order of about $10^{32}$ erg \citep{Emslie_etal_2012},
they are clearly different.
Flares are relatively local phenomena and the released energy is
mostly converted into radiation and energetic particles; while CMEs
are more global phenomena and the energy mostly goes into mechanical
energies through ejection of magnetized plasma structures. An intense
flare may not necessarily be accompanied by a CME \citep[e.g.,][]{Feynman_Hundhausen_1994, Green_etal_2002,
Yashiro_etal_2005, Wang_Zhang_2007}, because whether or not there is
a CME is substantially determined by the driving force of the inner
core magnetic field and the confining force of the external overlying
field \citep[e.g.,][]{Wang_Zhang_2007, LiuY_2008,
Schrijver_2009}. 

The inner driver is always in a form of highly sheared or twisted magnetic structure, e.g., a flux rope as required in most CME models \citep{Tamari_etal_1999, Torok_Kliem_2005}. Sheared or twisted field carries magnetic helicity, thus
provides a way to transport the helicity  naturally \citep{Low_1994, Tamari_etal_1999}. Since magnetic helicity is an invariant in the high conductive corona, it makes a point that a CME may be an inevitable product with the accumulation of helicity in corona \citep{Low_1994, Greenlm_etal_2002, Nindos_etal_2003, Zhang_2006, Zhang_flyer_low_2006, Zhang_Low_2008, Valori_etal_2012, Liu_Schuck_2012b}.

However, such an AR of continuously generating M and
X-class flares without a strong CME was rarely noticed before, and
particularly, 3 out of 6 non-CME X-class flares were of
long-duration (lasting more than one hour), which is quite
conflicting to many earlier studies that long-duration flares tend
to be easier to erupt out \citep[e.g.,][]{Harrison_1995,
Yashiro_etal_2006}. In order to understand the underlying physical
nature, we compare this super AR with other two pairs of ARs, 11157 and 11158, 11428 and 11429 firstly, then investigate the temporal evolution of photospheric parameters, pre-flare distribution of current helicity,  
and decay index of the five ARs in the next two sections. In the last section, we give the summary and the discussion.

\section{Activities of the ARs}

AR pairs 11157 and 11158 are chosen for comparison because their productivities are quite
different and they transited the visible solar disk almost during the same time, the same reason for AR 11428 and 11429.
 
AR 11157 was a very poor AR, no flare or CME was generated during its visible disk passage. In contrast, AR 11158 produced 68 flares above C1.0 and 12 CMEs, 10 of them were associated with flares above C1.0.
AR 11429, similar to AR 11158, was an prolific AR, too. It produced 64 flares above C1.0 and 16 CMEs, 10 were flare-accompanied. AR 11428 was chosen as a comparison for 11429, it produced only 5 C-class flares without any CME. It is noteworthy that AR 11429 and its neighbour AR 11430 were clustered, they had closely magnetic connection with each other,
and some weaker flares simultaneously occurred in AR 11430 when AR 11429 generated flares, the SHARP (Space weather HMI Active Region Patches, cutout maps of vector magnetograms that contain single AR (or cluster of ARs ) \citep{Bobra_etal_2014, Hoeksema_etal_2014} ) data also contains the two ARs. So we simply treat them as an AR complex here. Information of the ARs are showed in Table~\ref{tb_props}.
The associations of flares, CMEs and ARs are checked manually by using GOES 1.0--8.0 \AA\ soft X-ray flux data 
and the imaging data from SOHO/LASCO \citep{Brueckner_etal_1995}, STEREO/COR \citep{Kaiser_etal_2008} and SDO/AIA \citep{Lemen_etal_2012}.

\begin{figure*}
\begin{center}
\includegraphics[width=0.9\hsize]{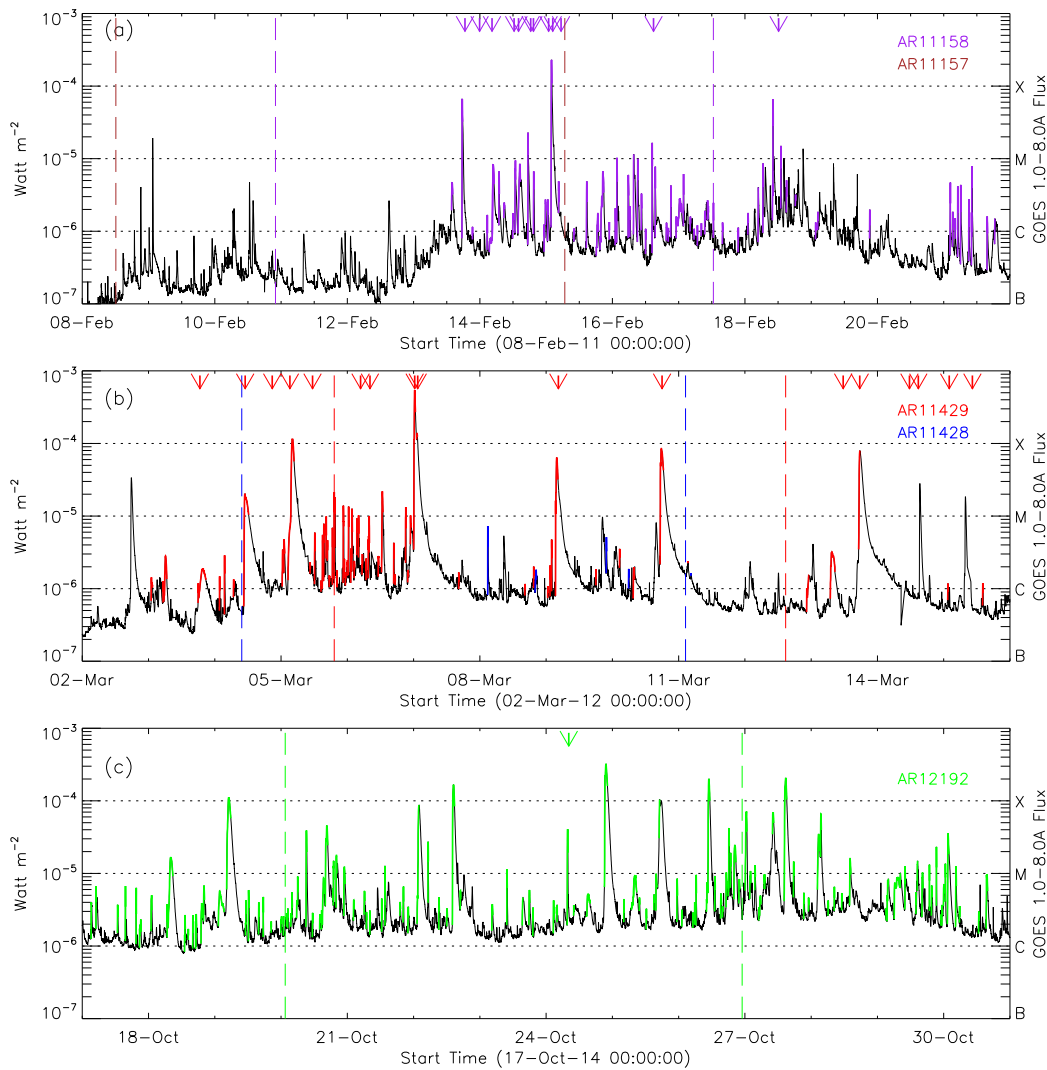}
\caption{GOES soft X-ray light curves during the ARs' disk passage, panel (a) for AR 11157 and 11158, (b) for AR 11428 and 11429, and (c) for AR 12192. Superimposed colored lines indicate the associated flares; brown, purple, blue, red and green colors are for AR 11157, 11158, 11428, 11429 and 12192, respectively. CMEs originating from the corresponding
ARs are marked by arrows. Vertical dashed lines indicate the time window when the central meridian distance (CMD) of the ARs' geometric centers were within $\pm 45^\circ$.} \label{fig:xray_allar}
\end{center}
\end{figure*}

For clarity, Fig.~\ref{fig:xray_allar} shows the GOES soft X-ray (SXR) flux during these ARs' transits.
AR 11157 and 11158 share the same light curve of the GOES SXR, same for AR 11428 and 11429. To distinguish the eruptions from the ARs, we use different colors for different ARs.
In the figure, all the associated flares are indicated by colored lines and the associated CMEs are marked by arrows. All the flares above M1.0 and the CMEs have been listed in Table~\ref{tb_acts}. The parameters of the flares are got from the SolarMonitor\footnote{\url{http://www.solarmonitor.org/}}, which is generated based on NOAA active region summaries, and the parameters of the CMEs are from the SOHO/LASCO CME Catalog\footnote{\url{http://cdaw.gsfc.nasa.gov/CME\_list/}} \citep{Yashiro_etal_2004} and the SECCHI/COR2 CME Catalog\footnote{\url{http://spaceweather.gmu.edu/seeds/secchi.php}} \citep{Olmedo_zhang_etal_2008}.

\section{Temporal evolution of the photospheric parameters}\label{sec_para}
 
Considering that an eruption, i.e., a flare or a CME, only lasts for a relatively short duration and an AR is not always active, 
we suggest that the overall features of the parameters during a noticeable duration may be 
more appropriate than instant values to characterize an AR's total productivity, 
though instant values are more valuable for the prediction of single event. Based on this idea, we carefully check the temporal evolution of all the SHARP parameters based on photospheric vector magnetograms \citep{Bobra_etal_2014}. The duration as indicated by the vertical dashed lines in Fig.~\ref{fig:xray_allar}, during which the central meridian distance (CMD) of the AR's geometric center is within $\pm 45^\circ$, is chosen to avoid the low signal-to-noise of the data near the solar limb.

\begin{figure*}
\begin{center}
\includegraphics[width=0.9\hsize]{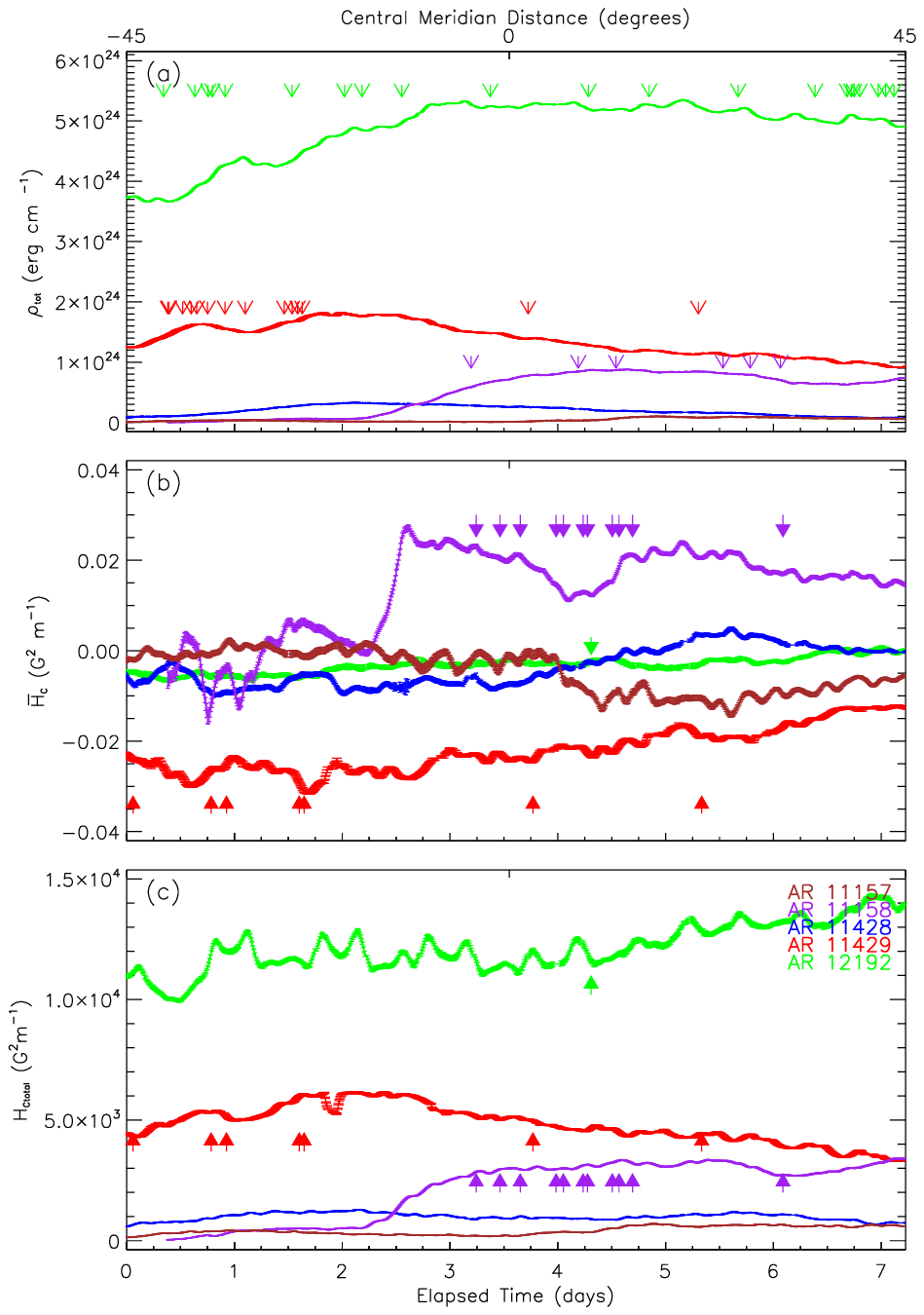}

\caption{The evolution of proxy of photospheric free magnetic energy  $\rho_{tot}$  (in panel (a)), mean current helicity $\overline{H_c}$ (in panel (b)) and total unsigned current helicity ${H_c}_{total}$  (in panel (c)); the hollow arrows in panel (a) indicate associated flares severer than M-class; the solid arrows in panel (b) and (c) are for CMEs; brown, purple, blue, red and green colors are for AR 11157, 11158, 11428, 11429 and 12192, respectively.} \label{fig:para}
\end{center}
\end{figure*}

We first find that the three parameters, the total magnetic flux ($\Phi$), total unsigned vertical current ($I_{total}$), proxy of photospheric free magnetic energy  ($\rho_{tot}$)
may be useful to distinguish ARs of different flare productivities. 
Since the evolutionary trends of them are similar, only $\rho_{tot}$ is showed in Fig.~\ref{fig:para}(a). The formal random error from the determination of vector field are overplotted as error bars, which are smaller compared 
to the values of the parameters themselves. Detailed formulas of the parameters are listed in Table~\ref{tb_paras}.

It is obvious that the two flare-rich ARs, 11429 and 12192 have larger $\rho_{tot}$ than the two inert ARs during 
their entire durations. The value of $\rho_{tot}$ of the other prolific AR, 11158, is small until it reached 
the central meridian since it's a newly emerged AR. There was no intense flare from the AR until $\rho_{tot}$ 
became large, as showed by the purple hollow arrows in Fig.~\ref{fig:para}(a). The 
mean values of $\Phi$, $I_{total}$ and $\rho_{tot}$ of AR 11158, 11429, 12192 during the duration, as showed in Table~\ref{tb_props}, are also larger than the two inert ones'. The results are consistent with the well-known picture that total flux is a physical measure of the size of an AR, reflecting the total magnetic energy contained in the AR; the electric current describes the deviation of the magnetic field from potential configuration, referring to the accumulation of free energy. The rapid dissipation of current could be manifested by flares, thus the strong current system seems to be a favorable condition for flares. 

It is hard to tell which parameter is more critical. As achieved by many previous studies, 
the combination of the three parameters is responsible for the flare productivity~\citep{Leka_Barnes_2007}. In fact, such a 
combination of the sharp parameters have been used for flare prediction: through employing a 
machine-learning algorithm, a set of quantities that mostly describe the magnetic energy and 
vertical current, could achieve a relatively high ability to distinguish flaring and non-flaring ARs~\citep{Bobra_Couvidat_2015}. 

We also find another parameter, mean current helicity ($\overline{H_c}$), being well consistent with the CME productivity, as showed in Fig.~\ref{fig:para} (b).
Though AR 12192 is the largest AR, 
its mean current helicity ($\overline{H_c}$)  is comparable to the two inert ARs', smaller than the two both-rich ARs'. 

Quantities describing current helicity are sometimes used as photospheric proxies of magnetic helicity 
\citep[e.g.,][]{Seehafer_1990, Abramenko_etal_1996, Bao_Zhang_1998,Zhang_Bao_1999, Zhang_etal_2000, Wang_Zhang_2015}. Magnetic helicity, defined as $H_m=\int A\cdot Bdx^3$, reflecting the twist, shear, linking or other non-potential properties of magnetic field, is approximately conserved in nearly-ideal MHD circumstance, e.g., the atmosphere of the sun \citep{Berger_Field_1984, Low_Berger_2003, Brown_etal_1999, Pevtsov_etal_1995, Demoulin_2007, Pevtsov_etal_2014}. Excess $H_m$ injected 
into the corona could be reorganized by locally resistant activities, finally erupt out in the form of highly sheared or twisted core field (e.g., flux ropes) contained in CMEs. It is pointed that ARs with eruptive flares contain more coronal magnetic helicity than the ones with confined flares \citep{ Nindos_Andrews_2004, Tziotziou_Georgoulis_2012}. In the nearly force-free state, the currents are almost parallel 
to the magnetic field lines, the helicity of current therefore could be a proxy of magnetic helicity. 

However, the mean current helicity ($\overline{H_c}$) that we used above is a signed average. Large value of $\overline{H_c}$ does indicate the presence of highly sheared or twisted field in one handedness, but small value of $\overline{H_c}$ could be the result of either absence of highly sheared or twisted field, or the offset between two highly sheared or twisted fields with opposite handedness.
Thus an additional parameter, total unsigned current helicity ${H_c}_{total}$, is checked to find the reason of small $\overline{H_c}$ of the three CME-poor ARs .  As showed in Fig.~\ref{fig:para} (c), AR 11157, 11428 have small ${H_c}_{total}$, but AR 12192 has quite large ${H_c}_{total}$, which means that there may be some sheared or twisted fields in AR 12192, too. Pre-existing, highly sheared or twisted core field could serve as
a seed structure for a CME, so why is AR 12192 still CME-poor? The position and the maturity of the core field and the confinement above the AR may be responsible. To further check this speculation,  we investigate the spatial distribution of current helicity on the photosphere 
and decay index of the magnetic field above the ARs in the next section.

\section{Pre-flare conditions}
\subsection{Spatial Distribution of Current Helicity}\label{subsec_hc}

We inspect the spatial distribution of current helicity $h_c$ on the photosphere of the ARs at specific moments: central meridian transits for the two inert ARs, 11157 and 11428; the moments right before the onsets of the largest flares of the three productive ARs: the X2.2 (2011-02-15T01:44) flare for 11158, the X5.4 (2012-03-07T00:02) flare for 11429, and the X3.1 (2014-10-24T21:07) flare for 12192. The former two flares are eruptive, and the last one is confined. 

\begin{figure*}
\begin{center}
\includegraphics[width=0.9\hsize]{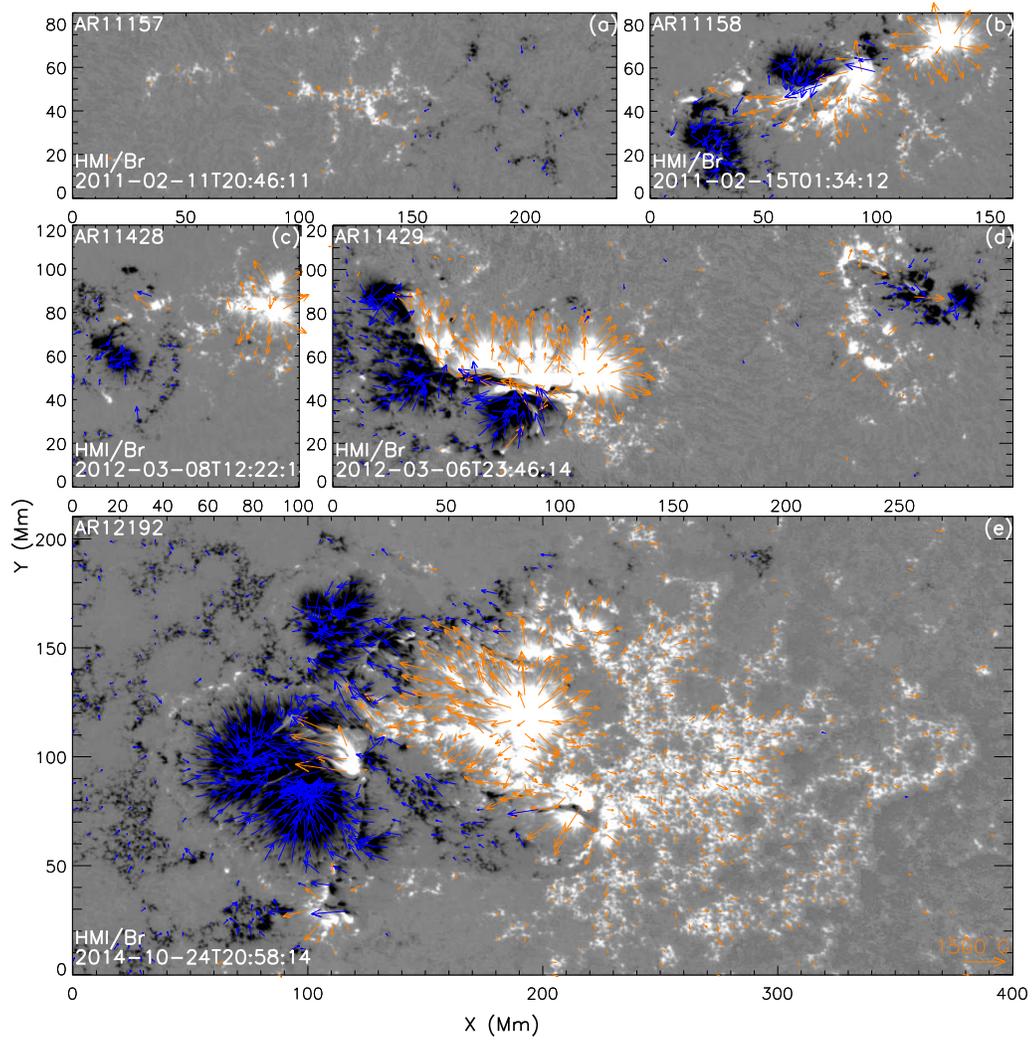}

\caption{Vector magnetic field at the specific moments of the five ARs, panels (a) to (e) are for AR 11157, 11158, 11428, 11429 and 12192, respectively. The background is $B_r$ component plotted in a dynamic range of $\pm1000$ Gausses, 
white (black) regions for the positive (negative) $B_r$. Orange (blue) arrows show the horizontal field component that originate from positive (negative) $B_r$ region.} \label{fig:field}
\end{center}
\end{figure*}

Fig.~\ref{fig:field} shows the vector magnetic field at the specific moments of the five ARs, $B_r$ component is plotted as background in a dynamic range of $\pm1000$ Gausses, 
white (black) patches for the positive (negative) $B_r$. Orange (blue) arrows show the horizontal field component, that originate from positive (negative) $B_r$ region. The panels are plotted in unit of Mm with the same scale, allowing direct comparison of the size of the ARs. Clealy AR 12192 is the largest one, having strong magnetic field as the other two productive ARs. It should be noted that there are some ``bad pixels" with abnormal weak $B_r$ in the center of the neagtive polarity of AR 12192, may be a result of failed inversion. We set thresholds on both the values of formal errors and the relative errors to the vector magnetic field, find out those pixels and smooth them with ambient pixels. Current helicity is calculated after the smooth. In each pixel, it is calculated by the formula $h_c = B_z \cdot  (\bigtriangledown \times B)_z ={\mu_0} B_z J_z  $, 
 in which the vertical current density $J_z$ is weighted by $B_z$, the vertical component of magnetic field,
which makes $h_c$ more sensitive to the twist or shear of vertical field under the force-free assumption.

\begin{figure*}
\begin{center}
\includegraphics[width=0.9\hsize]{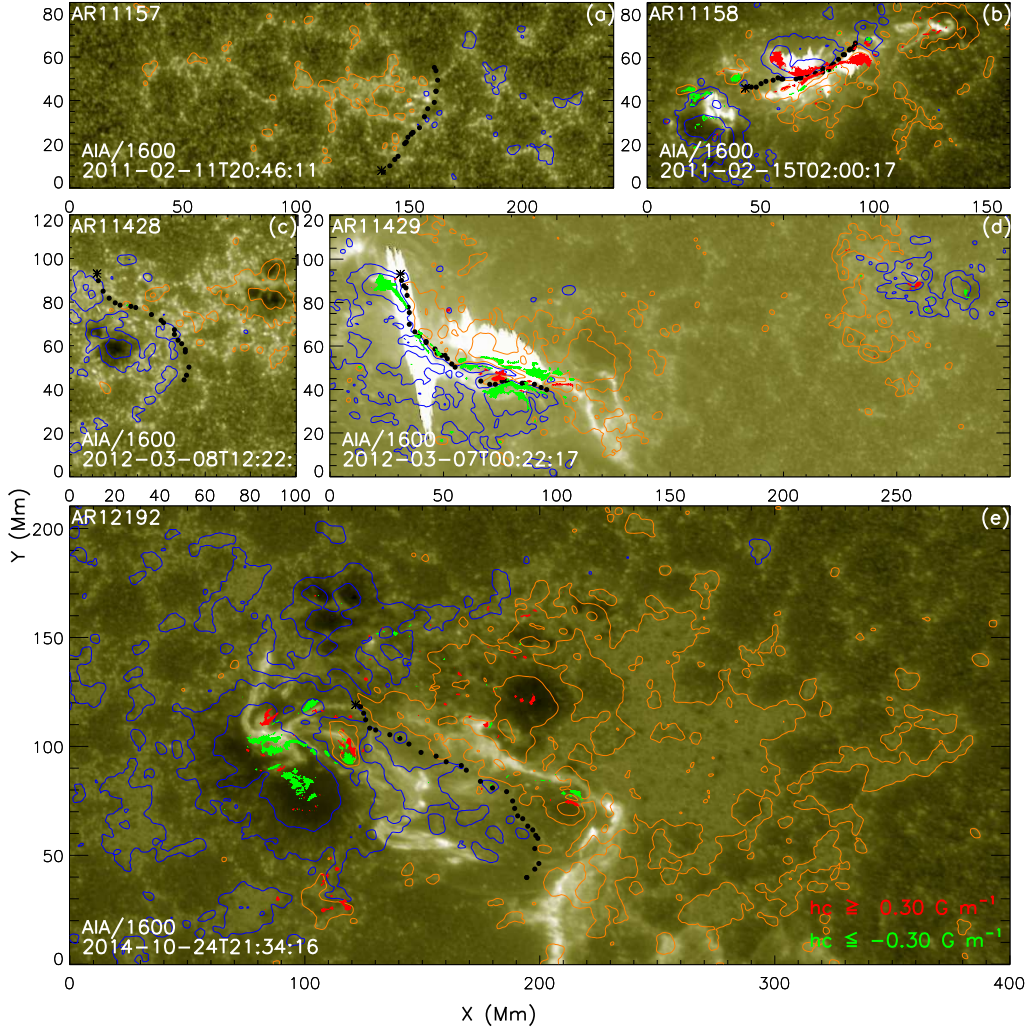}
\caption{Distribution of current helicity $h_c$ of the ARs at specific moments: right 
before the onsets of the biggest flares for AR 11158, 11429 and 12192; central meridian transits for 
11157 and 11428. $AIA/1600 \AA\ $ images near the flares' peak are plotted as background. $B_r$ component of the field is contoured on the images, orange contours for positive $B_r$ of $200,1000$ gausses, blue contours for negative $B_r$ of $-200,-1000$ gausses. Red patches are for $h_c \geqslant 0.3\ G^2m^{-1}$  and green ones for $h_c \leqslant -0.3\ G^2m^{-1}$. 
Black dotted lines show the paths along the flaring neutral lines or the main polarity inverse lines (if there was no flare), 
above which we calculate the decay index. The panels (a) to (e) are for AR 11157, 11158, 11428, 11429 and 12192, respectively. } \label{fig:hc}
\end{center}
\end{figure*}

Colored patches in Fig.~\ref{fig:hc} shows the distribution of the current helicity, red for the regions of $h_c \geqslant 0.3\ G^2 m^{-1}$ 
and green for $h_c \leqslant -0.3\ G^2 m^{-1} $, the threshold of $\pm0.3\ G^2 m^{-1}$ is about two times of standard deviation from mean value of $ h_c $ map of AR 12192, which could find regions of extremely large $ h_c $.   
The background are images of $AIA/1600 \AA\ $ near the flares' peak, showing the positions of the flare ribbons. $B_r$ component of the field is contoured on the images, orange contours for positive $B_r$ of $200,1000$ gausses , blue contours for negative $B_r$ of $-200,-1000$ gausses.

The black dotted lines indicate the neutral lines where the flares originated if any. 
One may expect the flare of AR 12192 mainly associated with the positive polarity within the major negative-polarity 
concentration, but the flare did light along the neutral line indicated 
by the black dotted line, as showed by the flare ribbons in Fig.~\ref{fig:hc}(e), that is also confirmed in Thalmann's work~\citep{Thalmann_etal_2015}. It is clear that in each of the two ARs with eruptive flares, AR 11158 and 11429, there is strong, concentrated current helicity along both sides of the neutral lines associated with the flare ribbons; while almost no strong current helicity exists in the two inert ARs, 11157 and 11428. In the AR with confined flare, AR 12192, strong, concentrated current helicity exists predominantly in the biggest negative polarity spot, far from the flaring neutral line.

The concentrated patches of strong current helicity along the flaring neutral line may indicate the photospheric footprints 
of the highly sheared or twisted core field that may serve as the seed of a CME. We think that for a mature core field, its field lines should come out from a strong $h_c$ region of the positive polarity, and go into a strong $h_c$ region of the negative polarity, which means that there should be strong $h_c$ regions in both positive and negative polarities on the photosphere. Then one may expect the magnetic flux in the positive and negative polarity patches with strong $h_c$ being roughly balanced. Thus we propose a ratio ($R^{\phi}$) between the magnetic flux contributed by the strong $h_c$ regions in both polarities to test this speculation. The current helicity and magnetic flux in the strong $h_c$ region of the three productive ARs are listed in Table.~\ref{tb_hcs}. For all pixels where $|h_c| \geqslant 0.3\ G^2m^{-1}$, $R^{\phi}$ of AR 11158, 11429 are not larger than 1.6, indicating a rough balance between the positive and negative magnetic flux in those regions; while $R^{\phi}$ of AR 12192 is 2.58, indicating a stronger flux imbalance. Furtherly, an AR usually has a dominant current helicity sign, as indicated by $\overline{H_c}$ in Fig.~\ref{fig:para} and Table.~\ref{tb_props}, positive for AR 11158, negative for AR 11429 and AR 12192. It may refer to a dominant  handedness of twist or shear in an AR. Thus we introduce another ratio, $R^\phi_d$, similar to $R^{\phi}$ but only for the pixels where $|h_c|$ greater than the threshold in the dominant sign. It is found that $R^\phi_d$ of AR 12192 is 4.67, much larger than the value of AR 11158 and AR 11429, showing a stronger flux imbalance of strong $h_c$ regions in both polarities. 
 These results suggest that there might be no mature sheared or twisted core field in AR 12192, 
and the large value of ${H_c}_{total}$ shown in Fig.~\ref{fig:para}c could be a result of the large area of the AR.  
For the two inert ARs, 11157 and 11428, there are no such seed structures, too, as exhibited by the $h_c$ distribution. 
The current helicity explored here can be easily derived from the measurements of the photosphere magnetic field, and could be a useful parameter for the space weather forecasting.

\subsection{Decay Index}\label{subsec_dec}

Further, the pre-existing core field may be constrained by the external field. 
We check the decay index of the ARs in this section to discover the constraint above the ARs.
Decay index is defined by $n=-\frac{d\ln B_{ex}(h)}{d\ln h}$, in which $h$ is the height from the solar surface, $B_{ex}$ is the external field above the AR. The coronal magnetic field here is potential field extrapolated from SDO/HMI synoptic chart by using the potential field source surface (PFSS) model \citep{Schatten_etal_1969, Wang_Sheeley_1992}. 
A larger decay index means that the constraint in the corona decrease faster with increasing height, and therefore 
a perturbation in the lower corona may cause the CME seed to erupt out more easily \citep{Torok_Kliem_2005, Wang_Zhang_2007, LiuY_2008}. Critical value above which an eruption is more likely to occur is $1.5$ 
 \citep{Torok_Kliem_2007, Aulanier_torok_2010}.

\begin{figure*}
\begin{center}
\includegraphics[width=\hsize]{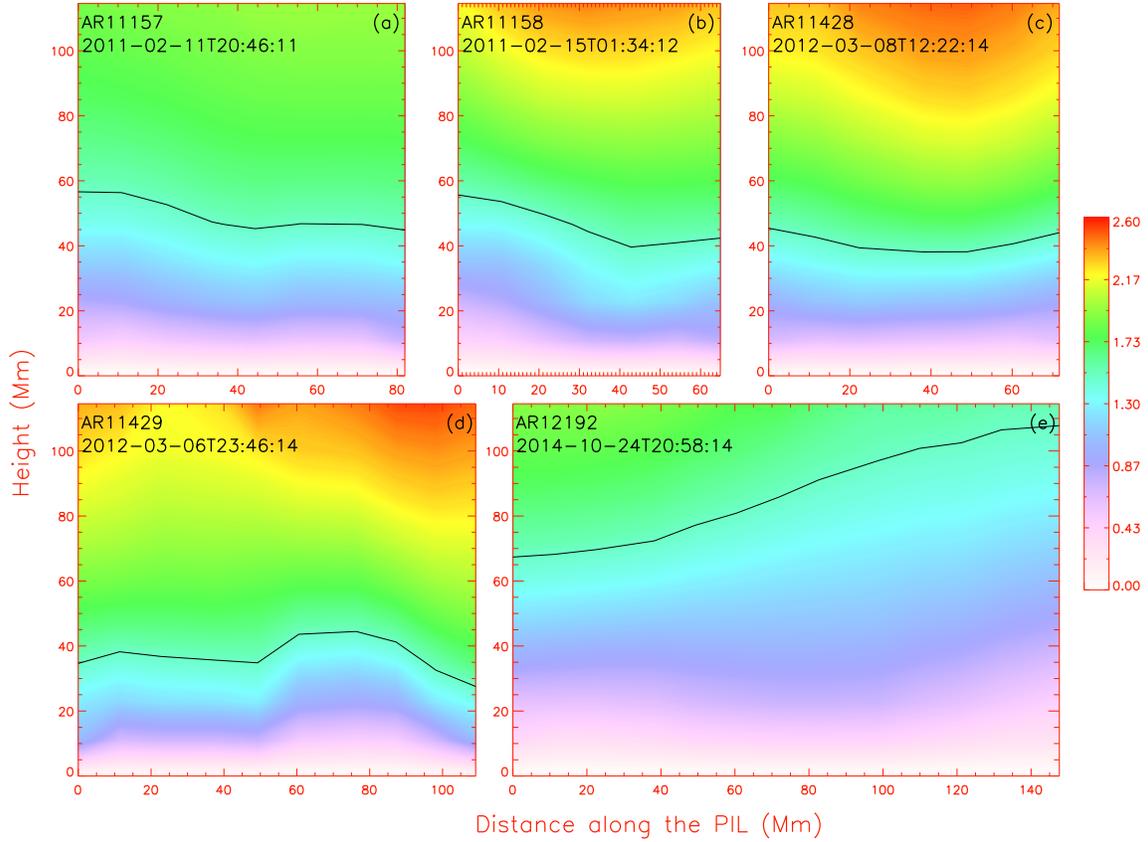}

\caption{Distribution of the decay index above the flaring neutral lines 
of the AR 11158, 11429 and 12192 before the onsets of their biggest flares (Panel b, d and e), 
or above the polarity inverse lines of the AR 11157 and 11428 
at their central meridian transits (Panel a and c). The black lines mark the position 
where $n$ reached the critical value of $1.5$.} \label{fig:decay}
\end{center}
\end{figure*}

Fig.~\ref{fig:decay} shows the spatial distribution of decay index along the flaring neutral lines, 
the black lines mark the critical heights where $n$ reaches $1.5$. Clearly, the heights of AR 11158 
and 11429 where $n$ reach 1.5 are lower than AR 12192, which means the constraining field above the 
two ARs with eruptive flares decays rapidly than the one with confined flare, making a CME more easily. 
The two inert ARs, 11157 and 11428 also have relatively low critical heights, but they have no
appreciable seed structures as pointed out before. Thus there was no CME even though the external field decayed rapidly.

\section{Concluding Remarks}\label{sec_con}

In this work, through comparing AR 12192 with other four ARs, we find that three parameters: 
the total magnetic flux ($\Phi$), total unsigned vertical current ($I_{total}$), 
proxy of photospheric free magnetic energy ($\rho_{tot}$), could be responsible for the flare productivity of 
our sample ARs. The flare-rich only AR 12192, same as the other two flare-rich ARs, 11158 and 
11429, has larger $\Phi$, $I_{total}$ and $\rho_{tot}$, which means that they have larger size, 
and contain stronger current system and more free magnetic energy than the two inert ARs 11157 and 11428. 
It is reasonable since sufficient amount of free magnetic energy is a necessary condition for an AR to power flares.%

No single threshold on any parameter could be used to distinguish the flare and CME productivity of the ARs,  but the combination of the mean current helicity and the total unsigned current helicity can be used to distinguish the flare and CME productivity. The magnitude of the mean current helicity ($|\overline{H_c}| $) is large for the CME-rich ARs, and small for AR 12192 and the other two CME-poor ARs, while the total unsigned current helicity (${H_c}_{total}$) of AR 12192 is as large as the two CME-rich ARs, indicating the presence of sheared or twisted field in all three flare-productive ARs. 
Considering the spatial distribution of current helicity, AR 12192 has $h_c$ concentrated in only one polarity, suggesting the absence of a mature seed structure for CME formation during flares. The CME-rich ARs can also be distinguished by the constraint of the overlying arcade field: AR 12192 has a smaller decay index than the CME-rich ARs, thus no strong CME accompanied the many intense flares it produced.

Our study here suggests that pre-existing seed structures at flaring position might be a necessary condition for CMEs. 
Besides, a large decay index above the AR's flaring neutral lines, which indicates a weak constraint, may be 
another necessary condition for CMEs. All these facts explain the unusual behaviour of the AR 12192: super flare-rich but CME-poor.
The conclusion is obtained based on a sample of five ARs, it's generality should be checked within a larger sample, which would be performed in the future.

\acknowledgments{We truly thank our anonymous referee for careful review and helpful comments that help us to revise this paper. We acknowledge the use of the data from HMI and AIA instruments
onboard Solar Dynamics Observatory (SDO), MDI and LASCO instruments
onboard Solar and Heliospheric Observatory (SOHO), and The
Geostationary Operational Environmental Satellite (GOES).
This work is supported by the grants from NSFC (41131065,
41421063, 41274173, 41574165, 41222031 and 41474151), CAS (Key Research
Program KZZD-EW-01-4), MOEC (20113402110001), MOST 973 key project (2011CB811403),
the fundamental research funds for the central universities, and the funds of the Thousand Young Talents Programme of China (R.L).}  

\bibliographystyle{agu}
\bibliography{Bib_e12}

\begin{thebibliography}{66}
\providecommand{\natexlab}[1]{#1}
\expandafter\ifx\csname urlstyle\endcsname\relax
  \providecommand{\doi}[1]{doi:\discretionary{}{}{}#1}\else
  \providecommand{\doi}{doi:\discretionary{}{}{}\begingroup
  \urlstyle{rm}\Url}\fi

\bibitem[{\textit{Abramenko et~al.}(1996)\textit{Abramenko, Wang, and
  Yurchishin}}]{Abramenko_etal_1996}
Abramenko, V.~I., T.~Wang, and V.~B. Yurchishin, {Analysis of electric current
  helicity in active regions on the basis of vector magnetograms}, \textit{Sol.
  Phys.}, \textit{168}, 75--89, \doi{10.1007/BF00145826}, 1996.

\bibitem[{\textit{Akiyama et~al.}(2007)\textit{Akiyama, Yashiro, and
  Gopalswamy}}]{Akiyama_etal_2007}
Akiyama, S., S.~Yashiro, and N.~Gopalswamy, {The CME-productivity associated
  with flares from two active regions}, \textit{Adv. Sp. Res.}, \textit{39}(9),
  1467--1470, \doi{10.1016/j.asr.2007.03.033}, 2007.

\bibitem[{\textit{Aulanier et~al.}(2010)\textit{Aulanier, T{\"{o}}r{\"{o}}k,
  D{\'{e}}moulin, and DeLuca}}]{Aulanier_torok_2010}
Aulanier, G., T.~T{\"{o}}r{\"{o}}k, P.~D{\'{e}}moulin, and E.~E. DeLuca,
  {Formation of Torus-Unstable Flux Ropes and Electric Currents in Erupting
  Sigmoids}, \textit{Astrophys. J.}, \textit{708}(1), 314--333,
  \doi{10.1088/0004-637X/708/1/314}, 2010.

\bibitem[{\textit{Bao and Zhang}(1998)}]{Bao_Zhang_1998}
Bao, S., and H.~Zhang, {Patterns of Current Helicity for Solar Cycle 22},
  \textit{Astrophys. J.}, \textit{496}, L43--L46, \doi{10.1086/311232}, 1998.

\bibitem[{\textit{Berger and Field}(1984)}]{Berger_Field_1984}
Berger, M.~a., and G.~B. Field, {The topological properties of magnetic
  helicity}, \textit{J. Fluid Mech.}, \textit{147}, 133--148,
  \doi{10.1017/S0022112084002019}, 1984.

\bibitem[{\textit{Bobra and Couvidat}(2015)}]{Bobra_Couvidat_2015}
Bobra, M.~G., and S.~Couvidat, {Solar Flare Prediction Using SDO /HMI Vector
  Magnetic Field Data With a Machine-Learning Algorithm}, \textit{Astrophys.
  J.}, \textit{798}(2), 135, \doi{10.1088/0004-637X/798/2/135}, 2015.

\bibitem[{\textit{Bobra et~al.}(2014)\textit{Bobra, Sun, Hoeksema, Turmon, Liu,
  Hayashi, Barnes, and Leka}}]{Bobra_etal_2014}
Bobra, M.~G., X.~Sun, J.~T. Hoeksema, M.~Turmon, Y.~Liu, K.~Hayashi, G.~Barnes,
  and K.~D. Leka, {The Helioseismic and Magnetic Imager (HMI) Vector Magnetic
  Field Pipeline: SHARPs – Space-Weather HMI Active Region Patches},
  \textit{Sol. Phys.}, \textit{289}(9), 3549--3578,
  \doi{10.1007/s11207-014-0529-3}, 2014.

\bibitem[{\textit{Brown et~al.}(1999)\textit{Brown, Canfield, and
  Pevtsov}}]{Brown_etal_1999}
Brown, M.~R., R.~C. Canfield, and A.~A. Pevtsov, {Magnetic helicity in space
  and laboratory plasmas}, \textit{Am. Geophys. Union, Geophys. Monogr. Ser.},
  \textit{111}, 1999.

\bibitem[{\textit{{Brueckner et~al}}(1995)}]{Brueckner_etal_1995}
{Brueckner, GE and Howard, RA and Koomen, MJ and Korendyke, CM and Michels, DJ
  and Moses, JD and Socker, DG and Dere, KP and Lamy, PL and Llebaria}, A.,
  {The Large Angle Spectroscopic Coronagraph (LASCO)}, \textit{Sol. Phys.},
  \textit{162}, 357--402, 1995.

\bibitem[{\textit{Canfield et~al.}(1999)\textit{Canfield, Hudson, and
  McKenzie}}]{Canfield_etal_1999}
Canfield, R.~C., H.~S. Hudson, and D.~E. McKenzie, {Sigmoidal morphology and
  eruptive solar activity}, \textit{Geophys. Res. Lett.}, \textit{26}(6),
  627--630, \doi{10.1029/1999GL900105}, 1999.

\bibitem[{\textit{Chen and Wang}(2012)}]{ChenA_Wang_2012}
Chen, A., and J.~Wang, {Quantifying solar superactive regions with vector
  magnetic field observations}, \textit{Astron. Astrophys.}, \textit{543}, 49,
  \doi{10.1051/0004-6361/201118037}, 2012.

\bibitem[{\textit{Chen et~al.}(2011)\textit{Chen, Wang, and
  Shen}}]{Chen_etal_2011}
Chen, C., Y.~Wang, and C.~Shen, {Statistical study of coronal mass ejection
  source locations: 2. Role of active regions in CME production}, \textit{J.
  Geophys. Res. Sp. Phys.}, \textit{116}(A12108), \doi{10.1029/2011JA016844},
  2011.

\bibitem[{\textit{D{\'{e}}moulin}(2007)}]{Demoulin_2007}
D{\'{e}}moulin, P., {Recent theoretical and observational developments in
  magnetic helicity studies}, \textit{Adv. Sp. Res.}, \textit{39}, 1674--1693,
  \doi{10.1016/j.asr.2006.12.037}, 2007.

\bibitem[{\textit{Emslie et~al.}(2012)\textit{Emslie, Dennis, Shih, Chamberlin,
  Mewaldt, Moore, Share, Vourlidas, and Welsch}}]{Emslie_etal_2012}
Emslie, a.~G., B.~R. Dennis, a.~Y. Shih, P.~C. Chamberlin, R.~a. Mewaldt, C.~S.
  Moore, G.~H. Share, A.~Vourlidas, and B.~T. Welsch, {Global Energetics of
  Thirty-Eight Large Solar Eruptive Events}, \textit{Astrophys. J.},
  \textit{759}(1), 71, \doi{10.1088/0004-637X/759/1/71}, 2012.

\bibitem[{\textit{Falconer et~al.}(2002)\textit{Falconer, Moore, and
  Gary}}]{Falconer_etal_2002}
Falconer, D.~A., R.~L. Moore, and G.~A. Gary, {Correlation of the Coronal Mass
  Ejection Productivity of Solar Active Regions with Measures of Their Global
  Nonpotentiality from Vector Magnetograms: Baseline Results},
  \textit{Astrophys. J.}, \textit{569}(2), 1016, 2002.

\bibitem[{\textit{Falconer et~al.}(2006)\textit{Falconer, Moore, and
  Gary}}]{Falconer_etal_2006}
Falconer, D.~A., R.~L. Moore, and G.~a. Gary, {Magnetic Causes of Solar Coronal
  Mass Ejections: Dominance of the Free Magnetic Energy over the Magnetic Twist
  Alone}, \textit{Astrophys. J.}, \textit{644}(2), 1258--1272,
  \doi{10.1086/503699}, 2006.

\bibitem[{\textit{Feynman and Hundhausen}(1994)}]{Feynman_Hundhausen_1994}
Feynman, J., and A.~J. Hundhausen, {Coronal mass ejections and major solar
  flares: The great active center of March 1989}, \textit{J. Geophys. Res. Sp.
  Phys.}, \textit{99}(A5), 8451--8464, 1994.

\bibitem[{\textit{Georgoulis and Rust}(2007)}]{Georgoulis_Rust_2007}
Georgoulis, M.~K., and D.~M. Rust, {Quantitative forecasting of major solar
  flares}, \textit{Astrophys. J. Lett.}, \textit{661}(1), L109, 2007.

\bibitem[{\textit{Green et~al.}(2002{\natexlab{a}})\textit{Green, {L{\'{o}}pez
  Fuentes}, Mandrini, D{\'{e}}moulin, {Van Driel-Gesztelyi}, and
  Culhane}}]{Greenlm_etal_2002}
Green, L.~M., M.~C. {L{\'{o}}pez Fuentes}, C.~H. Mandrini, P.~D{\'{e}}moulin,
  L.~{Van Driel-Gesztelyi}, and J.~L. Culhane, {The magnetic helicity budget of
  a CME-prolific active region}, \textit{Sol. Phys.}, \textit{208}, 43--68,
  \doi{10.1023/A:1019658520033}, 2002{\natexlab{a}}.

\bibitem[{\textit{Green et~al.}(2002{\natexlab{b}})\textit{Green, Matthews, van
  Driel-Gesztelyi, Harra, and Culhane}}]{Green_etal_2002}
Green, L.~M., S.~A. Matthews, L.~van Driel-Gesztelyi, L.~K. Harra, and J.~L.
  Culhane, {Multi-wavelength observations of an X-class flare without a coronal
  mass ejection.}, \textit{Sol. Phys.}, \textit{205}(2), 325--339,
  2002{\natexlab{b}}.

\bibitem[{\textit{Guo et~al.}(2007)\textit{Guo, Zhang, and
  Chumak}}]{Guo_etal_2007}
Guo, J., H.~Q. Zhang, and O.~V. Chumak, {Magnetic properties of flare-CME
  productive active regions and CME speed}, \textit{Astron. Astrophys.},
  \textit{462}(3), 1121--1126, \doi{10.1051/0004-6361:20065888}, 2007.

\bibitem[{\textit{Harrison}(1995)}]{Harrison_1995}
Harrison, R.~A., {The nature of solar flares associated with coronal mass
  ejection.}, \textit{Astron. Astrophys.}, \textit{304}, 585, 1995.

\bibitem[{\textit{Hoeksema et~al.}(2014)\textit{Hoeksema, Liu, Hayashi, and
  Sun}}]{Hoeksema_etal_2014}
Hoeksema, J., Y.~Liu, K.~Hayashi, and X.~Sun, {The Helioseismic and Magnetic
  Imager (HMI) Vector Magnetic Field Pipeline: Overview and Performance},
  \textit{Sol. Phys.}, \textit{289}, 3483--3530,
  \doi{10.1007/s11207-014-0516-8}, 2014.

\bibitem[{\textit{Jing et~al.}(2006)\textit{Jing, Song, and
  Abramenko}}]{Jing_etal_2006}
Jing, J., H.~Song, and V.~Abramenko, {The statistical relationship between the
  photospheric magnetic parameters and the flare productivity of active
  regions}, \textit{Astrophys. J.}, \textit{644}(2), 1273--1277, 2006.

\bibitem[{\textit{Kaiser et~al.}(2008)\textit{Kaiser, Kucera, Davila, Cyr,
  Guhathakurta, and Christian}}]{Kaiser_etal_2008}
Kaiser, M.~L., T.~a. Kucera, J.~M. Davila, O.~C.~S. Cyr, M.~Guhathakurta, and
  E.~Christian, {The STEREO mission: An introduction}, \textit{Space Sci.
  Rev.}, \textit{136}(1), 5--16, \doi{10.1007/s11214-007-9277-0}, 2008.

\bibitem[{\textit{Leka and Barnes}(2003)}]{Leka_Barnes_2003b}
Leka, K.~D., and G.~Barnes, {Photospheric Magnetic Field Properties of Flaring
  versus Flare‐quiet Active Regions. II. Discriminant Analysis},
  \textit{Astrophys. J.}, \textit{595}(2), 1296--1306, \doi{10.1086/377512},
  2003.

\bibitem[{\textit{Leka and Barnes}(2007)}]{Leka_Barnes_2007}
Leka, K.~D., and G.~Barnes, {Photospheric Magnetic Field Properties of Flaring
  versus Flare‐quiet Active Regions. IV. A Statistically Significant Sample},
  \textit{Astrophys. J.}, \textit{656}(2), 1173--1186, \doi{10.1086/510282},
  2007.

\bibitem[{\textit{Leka and Barnes.G}(2003)}]{Leka_Barnes_2003a}
Leka, K.~D., and Barnes.G, {Photospheric magnetic field properties of flaring
  versus flare-quiet active regions. i. data, general approach, and sample
  results}, \textit{Astrophys. J.}, \textit{595}, 1277--1295, 2003.

\bibitem[{\textit{Lemen et~al.}(2012)}]{Lemen_etal_2012}
Lemen, J.~R., et~al., {The Atmospheric Imaging Assembly (AIA) on the Solar
  Dynamics Observatory (SDO)}, \textit{Sol. Phys.}, \textit{275}(1-2), 17--40,
  \doi{10.1007/s11207-011-9776-8}, 2012.

\bibitem[{\textit{Lin and Forbes}(2000)}]{Lin_Forbes_2000}
Lin, J., and T.~G. Forbes, {Effects of reconnection on the coronal mass
  ejection process}, \textit{J. Geophys. Res. Sp. Phys.}, \textit{105}(A2),
  2375--2392, 2000.

\bibitem[{\textit{Liu}(2008)}]{LiuY_2008}
Liu, Y., {Magnetic Field Overlying Solar Eruption Regions and Kink and Torus
  Instabilities}, \textit{Astrophys. J. Lett.}, \textit{679}(2), L151, 2008.

\bibitem[{\textit{Liu and Schuck}(2012)}]{Liu_Schuck_2012b}
Liu, Y., and P.~W. Schuck, {Magnetic Energy and Helicity in Two Emerging Active
  Regions in the Sun}, \textit{Astrophys. J.}, \textit{761}(2), 105,
  \doi{10.1088/0004-637X/761/2/105}, 2012.

\bibitem[{\textit{Low}(1994)}]{Low_1994}
Low, B.~C., {Magnetohydrodynamic processes in the solar corona: Flares, coronal
  mass ejections, and magnetic helicity}, \textit{Phys. Plasmas},
  \textit{1}(5), 1684--1690, \doi{10.1063/1.870671}, 1994.

\bibitem[{\textit{Low and Berger}(2003)}]{Low_Berger_2003}
Low, B.~C., and M.~a. Berger, {A Morphological Study of Helical Coronal
  Magnetic Structures}, \textit{Astrophys. J.}, \textit{589}, 644--657,
  \doi{10.1086/374614}, 2003.

\bibitem[{\textit{M.D.Andrews and A.Nindos}(2004)}]{Nindos_Andrews_2004}
M.D.Andrews, and A.Nindos, {The association of big flares and coronal mass
  ejections :what is the role of magnetic helicity?}, \textit{Astrophys. J.},
  \textit{616}, 175--178, 2004.

\bibitem[{\textit{Nindos et~al.}(2003)\textit{Nindos, Zhang, and
  Zhang}}]{Nindos_etal_2003}
Nindos, A., J.~Zhang, and H.~Zhang, {The Magnetic Helicity Budget of Solar
  Active Regions and Coronal Mass Ejections}, \textit{Astrophys. J.},
  \textit{594}, 1033--1048, 2003.

\bibitem[{\textit{N.seehafer}(1990)}]{Seehafer_1990}
N.seehafer, {Electric current helicity in the solar atmosphere}, \textit{Sol.
  Phys.}, \textit{125}, 219--232, 1990.

\bibitem[{\textit{Olmedo et~al.}(2008)\textit{Olmedo, Zhang, Wechsler, Poland,
  and Borne}}]{Olmedo_zhang_etal_2008}
Olmedo, O., J.~Zhang, H.~Wechsler, A.~Poland, and K.~Borne, {Automatic
  detection and tracking of coronal mass ejections in coronagraph time series},
  \textit{Sol. Phys.}, \textit{248}(2), 485--499,
  \doi{10.1007/s11207-007-9104-5}, 2008.

\bibitem[{\textit{Pevtsov et~al.}(1995)\textit{Pevtsov, Canfield, and
  Metcalf}}]{Pevtsov_etal_1995}
Pevtsov, A.~a., R.~C. Canfield, and T.~R. Metcalf, {Latitudinal variation of
  helicity of photospheric magnetic fields}, \textit{Astrophys. J.},
  \textit{440}, 109--112, \doi{10.1086/187773}, 1995.

\bibitem[{\textit{Pevtsov et~al.}(2014)\textit{Pevtsov, Berger, Nindos, Norton,
  and van Driel-Gesztelyi}}]{Pevtsov_etal_2014}
Pevtsov, A.~a., M.~a. Berger, A.~Nindos, A.~a. Norton, and L.~van
  Driel-Gesztelyi, {Magnetic Helicity, Tilt, and Twist}, \textit{Space Sci.
  Rev.}, \textit{186}(1-4), 285--324, \doi{10.1007/s11214-014-0082-2}, 2014.

\bibitem[{\textit{RHESSI science nugget no.239}()\textit{RHESSI science nugget no.239}}]{RHESSI_Nugget_239}
{RHESSI} science nugget no.239,
  \url{http://sprg.ssl.berkeley.edu/~tohban/wiki/index.php/A_Record-Setting_CMEless_Flare}.

\bibitem[{\textit{Sammis et~al.}(2000)\textit{Sammis, Tang, and
  Zirin}}]{Sammis_etal_2000}
Sammis, I., F.~Tang, and H.~Zirin, {The dependence of large flare occurrence on
  the magnetic structure of sunspots}, \textit{Astrophys. J.}, \textit{540}(1),
  583, 2000.

\bibitem[{\textit{Schatten et~al.}(1969)\textit{Schatten, Wilcox, and
  Ness}}]{Schatten_etal_1969}
Schatten, K.~H., J.~M. Wilcox, and N.~F. Ness, {A model of interplanetary and
  coronal magnetic fields}, \textit{Sol. Phys.}, \textit{6}(3), 442--455, 1969.

\bibitem[{\textit{Schrijver}(2007)}]{Schrijver_2007}
Schrijver, C.~J., {A Characteristic Magnetic Field Pattern Associated with All
  Major Solar Flares and Its Use in Flare Forecasting}, \textit{Astrophys. J.},
  \textit{655}(2), L117--L120, \doi{10.1086/511857}, 2007.

\bibitem[{\textit{Schrijver}(2009)}]{Schrijver_2009}
Schrijver, C.~J., {Driving major solar flares and eruptions: A review},
  \textit{Adv. Sp. Res.}, \textit{43}(5), 739--755,
  \doi{10.1016/j.asr.2008.11.004}, 2009.

\bibitem[{\textit{Sun et~al.}(2015)}]{Sun_etal_2015}
Sun, X., et~al., {Why is the great solar active region 12192 flare-rich but
  cme-poor?}, \textit{Astrophys. J.}, \textit{804}, L28,
  \doi{10.1088/2041-8205/804/2/L28}, 2015.

\bibitem[{\textit{{T. Amari, J. F. Luciani, Z.
  Mikic}}(1999)}]{Tamari_etal_1999}
{T. Amari, J. F. Luciani, Z. Mikic}, J.~L., {Three-dimensional solutions of
  magnetohydrodynamic equations for prominence magnetic support: Twisted
  magnetic flux rope}, \textit{Astrophys. J.}, \textit{518}, 57--60, 1999.

\bibitem[{\textit{Ternullo et~al.}(2006)\textit{Ternullo, Contarino, Romano,
  and Zuccarello}}]{Ternullo_etal_2006}
Ternullo, M., L.~Contarino, P.~Romano, and F.~Zuccarello, {A statistical
  analysis of sunspot groups hosting M and X flares}, \textit{Astron.
  Nachrichten}, \textit{327}(1), 36--43, \doi{10.1002/asna.200510485}, 2006.

\bibitem[{\textit{Thalmann et~al.}(2015)\textit{Thalmann, Su, Temmer, and
  Veronig}}]{Thalmann_etal_2015}
Thalmann, J.~K., Y.~Su, M.~Temmer, and a.~M. Veronig, {The Confined X-Class
  Flares of Solar Active Region 2192}, \textit{Astrophys. J.}, \textit{801}(2),
  L23, \doi{10.1088/2041-8205/801/2/L23}, 2015.

\bibitem[{\textit{Tian et~al.}(2002)\textit{Tian, Liu, and
  Wang}}]{Tian_etal_2002a}
Tian, L., Y.~Liu, and J.~Wang, {The Most Violent Super-Active Regions in the
  22nd and 23rd Cycles}, \textit{Sol. Phys.}, \textit{209}(2), 361--374, 2002.

\bibitem[{\textit{Torok and Kliem}(2005)}]{Torok_Kliem_2005}
T{\"{o}}r{\"{o}}k, T., and B.~Kliem, {Confined and ejective eruptions of kink-unstable flux
  ropes}, \textit{Astrophys. J.}, \textit{630}, 97--100,
  \doi{10.1086/462412}, 2005.

\bibitem[{\textit{T{\"{o}}r{\"{o}}k and Kliem}(2007)}]{Torok_Kliem_2007}
T{\"{o}}r{\"{o}}k, T., and B.~Kliem, {Numerical simulations of fast and slow
  coronal mass ejections}, \textit{Astron. Nachrichten}, \textit{328}(8),
  743--747, \doi{10.1002/asna.200710795}, 2007.

\bibitem[{\textit{Tziotziou et~al.}(2012)\textit{Tziotziou, Georgoulis, and
  Raouafi}}]{Tziotziou_Georgoulis_2012}
Tziotziou, K., M.~K. Georgoulis, and N.-E. Raouafi, {The Magnetic
  Energy-Helicity Diagram of Solar Active Regions}, \textit{Astrophys. J.},
  \textit{759}(1), L4, \doi{10.1088/2041-8205/759/1/L4}, 2012.

\bibitem[{\textit{Valori et~al.}(2012)\textit{Valori, D{\'{e}}moulin, and
  Pariat}}]{Valori_etal_2012}
Valori, G., P.~D{\'{e}}moulin, and E.~Pariat, {Comparing values of the relative
  magnetic helicity in finite volumes}, \textit{Sol. Phys.}, \textit{278}(2),
  347--366, \doi{10.1007/s11207-012-9951-6}, 2012.

\bibitem[{\textit{Wang and Zhang}(2015)}]{Wang_Zhang_2015}
Wang, C., and M.~Zhang, {Correlation Between CME Occurrence Rate and Current
  Helicity in the Global Magnetic Field of Solar Cycle 23}, \textit{Sol.
  Phys.}, \textit{290}(46), 811--818, \doi{10.1007/s11207-015-0648-5}, 2015.

\bibitem[{\textit{Wang and Zhang}(2007)}]{Wang_Zhang_2007}
Wang, Y., and J.~Zhang, {A Comparative Study between Eruptive X-Class Flares
  Associated with Coronal Mass Ejections and Confined X-Class Flares},
  \textit{Astrophys. J.}, \textit{665}(2), 1428, 2007.

\bibitem[{\textit{Wang and Zhang}(2008)}]{Wang_Zhang_2008}
Wang, Y., and J.~Zhang, {A statistical study of solar active regions that
  produce extremely fast coronal mass ejections}, \textit{Astrophys. J.},
  \textit{680}(2), 1516--1522, 2008.

\bibitem[{\textit{Wang and {Sheeley Jr}}(1992)}]{Wang_Sheeley_1992}
Wang, Y.-M., and N.~R. {Sheeley Jr}, {On potential field models of the solar
  corona}, \textit{Astrophys. J.}, \textit{392}, 310--319, 1992.

\bibitem[{\textit{Yashiro}(2004)}]{Yashiro_etal_2004}
Yashiro, S., {A catalog of white light coronal mass ejections observed by the
  SOHO spacecraft}, \textit{J. Geophys. Res.}, \textit{109}(A7), A07,105,
  \doi{10.1029/2003JA010282}, 2004.

\bibitem[{\textit{Yashiro et~al.}(2005)\textit{Yashiro, Gopalswamy, Akiyama,
  Michalek, and Howard}}]{Yashiro_etal_2005}
Yashiro, S., N.~Gopalswamy, S.~Akiyama, G.~Michalek, and R.~A. Howard,
  {Visibility of coronal mass ejections as a function of flare location and
  intensity}, \textit{J. Geophys. Res. Sp. Phys.}, \textit{110}(A12S05),
  \doi{10.1029/2005JA011151}, 2005.

\bibitem[{\textit{Yashiro et~al.}(2006)\textit{Yashiro, Akiyama, Gopalswamy,
  and Howard}}]{Yashiro_etal_2006}
Yashiro, S., S.~Akiyama, N.~Gopalswamy, and R.~a. Howard, {Different Power-law
  Indices in the Frequency Distributions of Flares with and without Coronal
  Mass Ejections}, \textit{Astrophys. J.}, \textit{650}, 143--146,
  \doi{10.1086/508876}, 2006.

\bibitem[{\textit{Zhang and Bao}(1998)}]{Zhang_Bao_1999}
Zhang, H., and S.~Bao, {Latitudinal distribution of Photospheric Electric
  Current Helicity and Solar Activities}, \textit{Astron. Astrophys.},
  \textit{339}, 880--886, \doi{10.1086/307378}, 1998.

\bibitem[{\textit{Zhang et~al.}(2000)\textit{Zhang, Tian, Bao, and
  Zhang}}]{Zhang_etal_2000}
Zhang, H., L.~Tian, S.~Bao, and M.~Zhang, {Twist of Magnetic Fields in Solar
  Active Regions}, \textit{J. Astrophys. Astron.}, \textit{21}, 245--247,
  \doi{10.1134/1.163872}, 2000.

\bibitem[{\textit{Zhang}(2006)}]{Zhang_2006}
Zhang, M., {Helicity Observation of Weak and Strong Fields}, \textit{Astrophys.
  J.}, \textit{646}, 85--88, \doi{10.1086/506560}, 2006.

\bibitem[{\textit{Zhang and Flyer}(2008)}]{Zhang_Low_2008}
Zhang, M., and N.~Flyer, {The dependence of the helicity bound of force-free
  magnetic fields on boundary conditions}, \textit{Astrophys. J.},
  \textit{683}, 1160--1167, \doi{10.1086/589993}, 2008.

\bibitem[{\textit{Zhang et~al.}(2006)\textit{Zhang, Flyer, and
  Low}}]{Zhang_flyer_low_2006}
Zhang, M., N.~Flyer, and B.~C. Low, {Magnetic Field Confinement in the Corona:
  The Role of Magnetic Helicity Accumulation}, \textit{Astrophys. J.},
  \textit{644}(1), 575--586, \doi{10.1086/503353}, 2006.

\end{thebibliography}

\clearpage

\begin{table*}
\begin{center}
\footnotesize
\caption{Information and parameters of the ARs}\label{tb_props}
\begin{tabular}{l|ccccc}
\hline
NOAA & 11157 & 11158 & 11428 & 11429 & 12192 \\
\hline
Hemisphere & North & South & South & North & South \\
Date on the visible disk & 20110208-0217 & 20110211-0221 & 20120302-0313 & 20120303-0315 & 20141018-1030 \\
Productivity & Inert & Flare-CME-rich & Inert & Flare-CME-rich & Flare-rich only \\
$\Phi$ ($10^{22}Mx$) & $0.55  \pm 0.25 $ & $1.88  \pm 1.09 $ & $1.25  \pm 0.11 $ & $5.58  \pm 0.23 $ & $15.01 \pm 2.06 $ \\
$\rho_{tot}$ ($10^{23} erg\ cm^{-1}$) & $0.44  \pm 0.30$ & $4.97  \pm 3.34$ & $2.05  \pm 0.78$ & $13.88 \pm 2.64$ & $48.65 \pm 4.87$ \\
$I_{total}$ ($10^{13} A$) & $0.97  \pm 0.35$ & $4.08  \pm 2.19$ & $2.43  \pm 0.35$ & $9.45  \pm 1.20$ & $22.28 \pm 1.57$ \\
$\overline{H_c}$ ($10^{-3} G^2m^{-1}$) & $-4.62  \pm 4.69$ & $13.11  \pm 9.98$ & $-3.69  \pm 4.31$ & $-22.27 \pm 4.88$ & $-3.24  \pm 1.75$ \\ 
${H_c}_{total}$  ($10^{3} G^2m^{-1}$) & $0.42 \pm 0.17 $ & $2.19 \pm 1.20$ & $0.98 \pm 0.16$ & $4.76 \pm 0.81$ & $12.17 \pm 0.99$ \\

\hline    
\multicolumn{6}{l}{The lower 5 rows show the mean values and the standard deviations of the quantities during the period in which}\\
\multicolumn{6}{l}{the CMD was within $\pm 45^\circ$. See Table~\ref{tb_paras} for the formulas of the parameters.}
\end{tabular}\\
\end{center}
\end{table*}

\clearpage

\begin{longtable}{cc|ccccc|ccc}
\caption{ A table of the flares and CMEs from AR 11158, 11429 and 12192 $^*$} \label{tb_acts} \\ 
\hline
 &  & \multicolumn{5}{c|}{Flares} & \multicolumn{3}{c}{CMEs} \\
\hline
AR No. & No.  &  Date   & Begin &  End & Peak & Class & Time$^+$ & Width & Speed \\
 & & &(UT)   &  (UT)  &   (UT) &  & (UT) & (degree) & (km/s) \\
\hline
11158 & 1	& 2011/02/13 & 17:28:00 & 17:47:00 & 17:38:00 & M6.6	& 18:36:05 & 276 & 373\\
& 2 & 2011/02/13 & &&& &	23:54:00 & 73 & 370$^\dag$\\
& 3 	& 2011/02/14 & 04:29:00 & 05:09:00	&  04:49:00 &  C8.3 & 04:24:00 & 68  & 384$^\dag$\\
& 4 	& 2011/02/14 & 11:51:00 & 12:26:00	&  12:00:00 &  C1.7 & 12:24:00 & 61  & 810$^\dag$\\
& 5 	& 2011/02/14 & 13:47:00 & 14:42:00	&  14:27:00 &  C7.0 & 14:00:07 & 22  & 380\\
& 6 	& 2011/02/14 & 17:20:00 & 17:32:00	&  17:26:00 &  M2.2 & 18:24:05 & 360 & 326\\
& 7 	& 2011/02/14 & 19:23:00 & 19:36:00	&  19:30:00 &  C6.6 & 19:24:00 & 81  & 349$^\dag$\\
& 8 	& 2011/02/15 & 00:31:00 & 00:48:00	&  00:38:00 &  C2.7 & 00:54:00 & 82  & 1843$^\dag$\\
& 9 	& 2011/02/15 & 01:44:00 & 02:06:00	&  01:56:00 &  X2.2 & 02:24:05 & 360 & 669  	 \\
& 10	& 2011/02/15 & 04:27:00 & 04:37:00	&  04:32:00 &  C4.8 & 05:24:00 & 104 & 1748$^\dag$\\
& 11	& 2011/02/16 & 01:32:00 & 01:46:00  & 01:39:00  & M1.0	\\
& 12	& 2011/02/16 & 07:35:00 & 07:44:00  & 07:44:00  & M1.1	\\
& 13	& 2011/02/16 & 14:19:00 & 14:29:00	&  14:25:00 &  M1.6 & 14:54:00 & 75  & 320$^\dag$\\
& 14	& 2011/02/18 & 09:55:00 & 10:15:00  & 10:11:00  & M6.6	\\
& 15  & 2011/02/18 & & & & & 12:12:05 & 89 & 350\\
& 16	& 2011/02/18 & 12:59:00 & 13:06:00 & 13:03:00 & M1.4	&\\

\hline
11429 & 1  & 2012/03/03 & 17:56:00 & 18:05:00 & 18:03:00 & C1.1 & 18:36:05 & 192 & 1078 \\
  & 2  & 2012/03/04 & 10:29:00 & 12:16:00 & 10:52:00 & M2.0 & 11:00:07 & 360 & 1306 \\
  & 3  & 2012/03/04 &          &          &          &      & 20:48:05 & 50  &720\\
  & 4  & 2012/03/05 & 02:30:00 & 04:43:00 & 04:05:00 & X1.1 & 03:12:09 & 92  &594\\
  & 5  & 2012/03/05 &          &          &          &      & 11:24:06 & 50  &428 \\
  & 6  & 2012/03/05 & 19:10:00 & 19:21:00 & 19:16:00 & M2.1 &\\       
  & 7  & 2012/03/05 & 19:27:00 & 19:32:00 & 19:30:00 & M1.8 &\\       
  & 8  & 2012/03/05 & 22:26:00 & 22:42:00 & 22:34:00 & M1.3 &\\       
  & 9  & 2012/03/06 & 00:22:00 & 00:31:00 & 00:28:00 & M1.3 &\\       
  & 10 & 2012/03/06 & 01:36:00 & 01:50:00 & 01:44:00 & M1.2 &\\       
  & 11 & 2012/03/06 & 04:01:00 & 04:08:00 & 04:05:00 & M1.0 & 04:48:06 & 111 & 536 \\
  & 12 & 2012/03/06 & 07:52:00 & 08:00:00 & 07:55:00 & M1.0 & 08:12:08 & 107 & 599 \\
  & 13 & 2012/03/06 & 12:23:00 & 12:54:00 & 12:41:00 & M2.1 &\\       
  & 14 & 2012/03/06 & 21:04:00 & 21:14:00 & 21:11:00 & M1.3 &\\       
  & 15 & 2012/03/06 & 22:49:00 & 23:11:00 & 22:53:00 & M1.0 &\\       
  & 16 & 2012/03/07 & 00:02:00 & 00:40:00 & 00:24:00 & X5.4 & 00:24:06 & 360 & 2684 \\
  & 17 & 2012/03/07 & 01:05:00 & 01:23:00 & 01:14:00 & X1.3 & 01:30:24 & 360 & 1825 \\
  & 18 & 2012/03/09 & 03:22:00 & 04:18:00 & 03:53:00 & M6.3 & 04:26:09 & 360 & 950 \\
  & 19 & 2012/03/10 & 17:15:00 & 18:30:00 & 17:46:00 & M8.4 & 18:00:05 & 88  & 491 \\
  & 20 & 2012/03/13 &          &          &          &      & 11:36:05 & 54  &229 \\
  & 21 & 2012/03/13 & 17:12:00 & 17:41:00 & 17:30:00 & M7.9 & 17:36:05 & 360 & 1884\\
  & 22 & 2012/03/14 &          &          &          &      & 11:36:05 & 11  &565 \\
  & 23 & 2012/03/14 &          &          &          &      & 14:48:05 & 28  &533 \\
  & 24 & 2012/03/15 &          &          &          &      & 02:00:05 & 74  &318 \\
  & 25 & 2012/03/15 &          &          &          &      & 10:24:05 & 31  &388 \\
\hline
{12192} & 1  &  2014/10/18 & 07:02:00 & 08:49:00 & 07:58:00 &M1.6 &\\
  & 2   &2014/10/19 & 04:17:00 & 05:48:00 & 05:03:00 & X1.1 &\\
  & 3   &2014/10/20 & 09:00:00 & 09:20:00 & 09:11:00 & M3.9 &\\
  & 4   &2014/10/20 & 16:00:00 & 16:55:00 & 16:37:00 & M4.5 &\\
  & 5   &2014/10/20 & 18:55:00 & 19:04:00 & 19:02:00 & M1.4 &\\
  & 6   &2014/10/20 & 19:53:00 & 20:13:00 & 20:03:00 & M1.7 &\\
  & 7   &2014/10/20 & 22:43:00 & 23:13:00 & 22:55:00 & M1.2 &\\
  & 8   &2014/10/21 & 13:35:00 & 13:40:00 & 13:38:00 & M1.2 &\\
  & 9   &2014/10/22 & 01:16:00 & 01:59:00 & 01:59:00 & M8.7 &\\
  & 10  &2014/10/22 & 05:11:00 & 05:21:00 & 05:17:00 & M2.7 &\\
  & 11  &2014/10/22 & 14:02:00 & 14:50:00 & 14:28:00 & X1.6 &\\
  & 12  &2014/10/23 & 09:44:00 & 09:56:00 & 09:50:00 & M1.1 &\\
  & 13  &2014/10/24 & 07:37:00 & 07:53:00 & 07:48:00 & M4.0 & 08:00:05 & 96 & 677 \\
  & 14  &2014/10/24 & 21:07:00 & 22:13:00 & 21:40:00 & X3.1 &\\
  & 15  &2014/10/25 & 16:55:00 & 17:08:00 & 17:08:00 & X1.0 &\\
  & 16  &2014/10/26 & 10:04:00 & 11:18:00 & 10:56:00 & X2.0 &\\
  & 17  &2014/10/26 & 17:08:00 & 17:30:00 & 17:17:00 & M1.0 &\\
  & 18  &2014/10/26 & 18:07:00 & 18:20:00 & 18:15:00 & M4.2 &\\
  & 19  &2014/10/26 & 18:43:00 & 18:56:00 & 18:49:00 & M1.9 &\\
  & 20  &2014/10/26 & 19:59:00 & 20:45:00 & 20:21:00 & M2.4 &\\
  & 21  &2014/10/27 & 00:06:00 & 00:44:00 & 00:34:00 & M7.1 &\\
  & 22  &2014/10/27 & 01:44:00 & 02:11:00 & 02:02:00 & M1.0 &\\
  & 23  &2014/10/27 & 03:35:00 & 03:48:00 & 03:41:00 & M1.3 &\\
  & 24  &2014/10/27 & 09:59:00 & 10:26:00 & 10:09:00 & M6.7 &\\
  & 25  &2014/10/27 & 14:12:00 & 15:09:00 & 14:47:00 & X2.0 &\\
  & 26  &2014/10/27 & 17:33:00 & 17:47:00 & 17:40:00 & M1.4 &\\
  & 27  &2014/10/28 & 02:15:00 & 03:08:00 & 02:41:00 & M3.4 &\\
  & 28  &2014/10/28 & 03:23:00 & 03:41:00 & 03:32:00 & M6.6 &\\
  & 29  &2014/10/28 & 13:54:00 & 14:23:00 & 14:06:00 & M1.6 &\\
  & 30  &2014/10/29 & 06:03:00 & 08:20:00 & 08:20:00 & M1.0 &\\
  & 31  &2014/10/29 & 09:54:00 & 10:06:00 & 10:01:00 & M1.2 &\\
  & 32  &2014/10/29 & 14:24:00 & 14:33:00 & 14:33:00 & M1.4 &\\
  & 33  &2014/10/29 & 16:06:00 & 16:33:00 & 16:20:00 & M1.0 &\\
  & 34  &2014/10/29 & 18:47:00 & 18:52:00 & 18:50:00 & M1.3 &\\
  & 35  &2014/10/29 & 21:18:00 & 21:25:00 & 21:22:00 & M2.3 &\\
  & 36  &2014/10/30 & 00:34:00 & 00:40:00 & 00:37:00 & M1.3 &\\
  & 37  &2014/10/30 & 01:19:00 & 01:56:00 & 01:35:00 & M3.5 &\\
  & 38  &2014/10/30 & 04:17:00 & 04:36:00 & 04:28:00 & M1.2 &\\
\hline
\end{longtable}
$^*$ Blanks in the flare (CME) column mean no C-class or intenser flare (CME) associated with the CME (flare).
$^+$ First appearance in the field of view of the SOHO/LASCO C2 or the STEREO/COR2 (Missed by C2).
$^\dag$ CMEs recorded by COR2.

\begin{table*}
\begin{center}
\footnotesize
\caption{Parameters used to distinguish the AR's productivity$^*$}\label{tb_paras}
\begin{tabular}{ccccc}
\hline
Parameters & Description & Unit & Formula & Statistic \\
\hline    
$\Phi$  & Total unsigned flux             & $Mx$             & $\Phi=\Sigma|B_z|dA $  & Integral \\
$\rho_{tot}$  & Proxy for total photospheric     & $erg \ cm^{-1}$  & $\rho_{tot} = \Sigma \frac{1}{8 \pi} (B^{obs}-B^{Pot})^2dA $  & Integral \\
        & excess magnetic energy  &                  &   \\
$I_{total}$ & total unsigned vertical current & $A$              & $I_{total}=\Sigma|J_z|dA$  & Integral \\
$ \overline{H_c} $ & mean current helicity ($B_z$ contribution)  & $G^2 m^{-1}$     & $ \overline{H_c} = \frac{1 }{N} \Sigma B_z (\bigtriangledown \times B)_z $ & Mean \\
${H_c}_{total}$ & Total unsigned current helicity  & $G^2 m^{-1}$     & ${H_c}_{total} = {\Sigma | B_z (\bigtriangledown \times B)_z | } $ & Sum \\
\hline
\end{tabular}\\
$^*$ Adapted from \citet{Bobra_etal_2014}. Here ${\mu_0} J_z= (\bigtriangledown \times B)_z= (\frac{\partial B_y}{\partial x}-\frac{\partial B_x}{\partial y})$
\end{center}
\end{table*}

\begin{table*}
\begin{center}
\footnotesize
\caption{Current helicity and magnetic flux in strong $h_c$ regions in different polarities$^*$}\label{tb_hcs}
\begin{tabular}{c|cc|cc|c|cc|cc|c}
\hline
AR No. & \multicolumn{4}{c}{Parameters in all strong $h_c$ pixels} &  & \multicolumn{5}{c}{Parameters in strong $h_c$ pixels of dominant sign} \\
\hline  
 & \multicolumn{2}{c|}{${H^t_c}_{total}$ ($G^2m^{-1}$) }  & \multicolumn{2}{c|}{$\Phi^t$ ($10^{21} Mx$)} & $R^{\Phi}$  & \multicolumn{2}{c|}{${{H^t_d}_c}$ ($G^2m^{-1}$)}  & \multicolumn{2}{c|}{$\Phi_d^t$ ($10^{21} Mx$) }  & $R_d^{\Phi}$ \\
&in $\mathcal{P}_{B_z>0}$ &in $\mathcal{P}_{B_z<0}$ &in $\mathcal{P}_{B_z>0}$ & in $\mathcal{P}_{B_z<0}$ &   &in $\mathcal{P}_{B_z>0}$ &in $\mathcal{P}_{B_z<0}$ &in $\mathcal{P}_{B_z>0}$ & in $\mathcal{P}_{B_z<0}$ & \\ 
\hline

11158 &  885.70 & 674.69 & 2.11 & -2.24 & 1.06 & 625.86 & 441.28 & 1.64 & -1.46 & 1.12\\
11429 & 964.28 & 1642.08 & 2.83 & -4.47 & 1.58 & -732.27 & -1381.21 & 2.24 & -3.83 & 1.71 \\
12192 & 1307.59 & 1988.95 & 3.58 & -9.23 & 2.58  &  -589.88 & -1385.31 & 1.49 & -6.95 & 4.67\\

\hline
\end{tabular}\\
\end{center}

$^*$ $\mathcal{P}_{B_z > 0}$($\mathcal{P}_{B_z <0}$) refers to positive (negative) polarity.\\
Strong $h_c$ regions refer to pixels where $|h_c| \geqslant 0.3\ G^2 m^{-1}$\\
Quantities are calculated by:\\
${H^t_c}_{total} = {\Sigma^t | B_z (\bigtriangledown \times B)_z | }$ ; ${{H^t_d}_c} = {\Sigma^t_d  B_z (\bigtriangledown \times B)_z  }$;
$\Phi^t=\Sigma^t B_z dA $; 
$\Phi_d^t=\Sigma^t_d B_z dA $\\
$R^{\Phi}$($R_d^{\Phi}$) is the larger ratio between $|\Phi^t|$($|\Phi_d^t|$) in different polarities.\\
Superscript $^t$ means threshold ($0.3\ G^2 m^{-1}$ here) of $h_c$; subscript $_d$ means dominant sign of $h_c$.\\

\end{table*}

\clearpage

\end{document}